\documentclass[fleqn,10pt]{wlscirep}

\usepackage{multirow,amssymb,amsbsy,amsmath,epstopdf}

\newcommand{\braket}[2]{\langle #1#2\rangle}

\newcommand{\ita}[1]{\textit{#1}}

\title{Complex quantum networks as structured environments: engineering and probing}

\author[1,*]{Johannes Nokkala}
\author[2]{Fernando Galve}
\author[2]{Roberta Zambrini}
\author[1]{Sabrina Maniscalco}
\author[1]{Jyrki Piilo}
\affil[1]{Turku Centre for Quantum Physics, Department of Physics and Astronomy,
University of Turku, FI-20014, Turun Yliopisto, Finland}
\affil[2]{IFISC (UIB-CSIC), Instituto de Fisica Interdisciplinary Sistemas Complejos, UIB Campus, 07122 Palma de Mallorca, Spain}

\affil[*]{jsinok@utu.fi}

\begin{abstract}
We consider structured environments modeled by bosonic quantum networks and investigate the probing of their spectral density, structure, and topology. We demonstrate how to engineer a desired spectral density by changing the network structure. Our results show that the spectral density can be very accurately detected via a locally immersed quantum probe for virtually any network configuration. Moreover, we show how the entire network structure can be reconstructed by using a single quantum probe. We illustrate our findings presenting examples of spectral densities and topology probing for networks of genuine complexity.
\end{abstract}

\begin{document}

%\flushbottom
\maketitle

\section*{Introduction}

Recently, the study of dissipative or decohering environments as engineered reservoirs has received considerable  attention. Reservoir engineering aims at turning the environment from an enemy to an ally by modifying the properties of the unavoidable noise in order to optimally preserve quantum features such as coherence or entanglement. It is nowadays generally recognised that the initial belief of the environment as the major enemy of all quantum technologies must be reconsidered. Theoretical and experimental results show indeed that the interaction with the environment, when suitably tailored, can be used to  generate or protect quantum resources, and holds great potential as a method to increase the feasibility and scale of quantum applications \cite{VWC, multiparticleendyn, DDengineer, engen, scirepsynchro}. 

On the other hand, the extraction of relevant information about a complex quantum system by means of a localised quantum probe has also enjoyed growing attention. Besides being generally a less invasive technique, this approach might significantly reduce the resources and effort needed to detect, e.g., the complex system's temperature \cite{probeT}, communities \cite{probeStruct}, state \cite{probeState} or dimensionality \cite{probeNM}.

In this work we apply concepts and techniques of open quantum systems theory to investigate complex bosonic quantum networks, thus bridging the gap between these two fields. More precisely, we consider a bosonic probe system locally immersed in the network and investigate the control and probing of the network spectral density $J(\omega)$, a quantity embedding both the environment structure and the interaction between system and environment \cite{Jwrules,OQS}. Notably, we demonstrate that not only $J(\omega)$ but the entire structure and topology of the network can be obtained by quantum probing, and we give a detailed method to accomplish this task. We consider the probe as an open quantum system whose dynamics arises from the interaction with a nontrivial quantum environment.  

The study of the spectral density of bosonic quantum systems has been addressed before only for the simple cases of either non-interacting oscillators or oscillator chains \cite{Rubin, dimerchainpaper, linearmappingA, linearmappingB, linearmappingC}. This leaves the determination and especially control of spectral densities of complex quantum networks a largely unexplored area. Previous investigations include using the Landau-Zener transition probabilities of a qubit to detect global features of $J(\omega)$  \cite{Earlier2006}, the center of mass motion of a michromechanical oscillator in the high temperature limit  \cite{Earlier2013}, the Stokes shift response function of a spin system immersed in a proteomic scaffold and in contact to a solvent \cite{Earlier2014} and the statistics of photons scattering from an open quantum system coupled to a zero temperature reservoir \cite{Earlier2014b}. In comparison, our probing scheme has the advantage of being local and of making no assumptions about the temperature or the structure of the environment. Networks of spins as well as bosonic/fermionic networks have been also investigated in Refs. \cite{Daniel2009,Daniel2011,Kato2014} in the context of structure probing. However, these methods require a known topology and full state tomography to detect the coupling and local field strengths \cite{Daniel2009,Daniel2011},  or can probabilistically reveal an unknown topology but are limited to small networks only \cite{Kato2014}. On the contrary, our approach is not based on state tomography, requires minimal post-processing of measured data, is in principle deterministic, and can be used for complex quantum networks having hundreds of nodes.

After introducing the model in detail, we calculate spectral densities of quantum networks of genuine complexity, e.g., small-world \cite{smallworld} and Erd\H{o}s-R\'{e}nyi \cite{erdosrenyi} configurations, and show how to control some features of their structure, such as the number of gaps. Numerical simulations based on exact diagonalization show that the scheme is very accurate for virtually any network configuration and temperature. While probing is done in the weak coupling regime, the detected $J(\omega)$ can be scaled to any coupling strength once its shape is known. Finally, we prove that the whole structure and topology of the network can be obtained with a single quantum probe by scanning over the network nodes with single and pair-node couplings.

\section*{Results}

{\textbf{The microscopic model.}} The probe is a quantum harmonic oscillator of unit mass coupled linearly with coupling strength $k$ to a node in the quantum network. The probe Hamiltonian, in units of $\hbar$, is $H_S=(p_{S}^2+\omega_{S}^2q_{S}^2)/2$, where $p_{S}$ and $q_{S}$ are the probe momentum and position operators and $\omega_{S}$ is the probe frequency. The network plays the role of a finite environment in a thermal state of temperature $T$ and consists of $N$ quantum harmonic oscillators of unit mass coupled via springlike couplings, each having the same bare frequency $\omega_0$. We assume that the probe and the network are initially uncorrelated.

The Hamiltonian of a generic oscillator network can be given as $H_{E}=\textbf{p}^{T}\textbf{p}/2+\textbf{q}^{T}\textbf{Aq}$, where $\textbf{p}=\left\lbrace p_{1}, p_{2}, ..., p_{N}\right\rbrace ^T  $ and $\textbf{q}=\left\lbrace q_{1}, q_{2}, ..., q_{N}\right\rbrace ^T  $ are the vectors of momentum and position operators and the $ N\times N $ matrix \textbf{A} has elements $ A_{ij}=\delta_{ij}\omega_i^2/2 - (1 - \delta_{ij})h_{ij}/2 $, where $ \omega_i $ is the effective frequency of oscillator $ i $, and $h_{ij}$ is the strength of the springlike coupling between oscillators $i$ and $j$. Matrix $\textbf{A}$ is the adjacency matrix (including diagonal terms) of the network and as such completely characterizes it. For any network configuration, the network Hamiltonian is quadratic in position and momentum and can therefore be diagonalized with an orthogonal transformation \textbf{K} as $\textbf{K}^T \textbf{AK} = \textbf{D} $, where the diagonal matrix $\bf{D}$ has elements $D_{ii}=\Omega_i^2/2$, where $\Omega_i$ are the eigenfrequencies of the network.

By defining new variables $\textbf{P}=\textbf{K}^T \textbf{p}$ and $\textbf{Q}=\textbf{K}^T \textbf{q}$, the diagonalization allows us to move into an equivalent eigenmode picture of independent oscillators, each interacting with the probe with a coupling strength directly proportional to $k$. The network Hamiltonian in the eigenmode picture is of the form $H_E=\sum\nolimits_{i=1}^N (P_{i}^2+\Omega_{i}^2Q_{i}^2)/2$, where $P_{i}$ and $Q_{i}$ are the position and momentum operators of the network eigenmodes which have frequencies $\Omega_i$. The Hamiltonian of the total system is $H=H_S+H_E+kH_I$, where $H_I$ is the interaction Hamiltonian. In the eigenmode picture it is of the form $H_I=q_S\sum\nolimits_{i=1}^N g_iQ_i$ where the dimensionless constants $g_i$ describe the coupling strengths between the probe and the eigenmodes and are just elements of a row of \textbf{K}. This can be seen by writing the position operator of the node directly coupled to the probe in terms of $Q_i$ and identifying the weights given by the elements of \textbf{K} with $g_i$.  Further details about oscillator network diagonalization can be found in, e.g., Appendix A of reference \cite{dimerchainpaper} by Vasile et al.

%%%%%%%%%%%% FIGURE 1 %%%%%%%%%%%%%%%%%%
\begin{figure}[t]
\centerline{
	\includegraphics[trim=0cm 0cm 0cm 0cm,clip=true,width=0.95\textwidth]{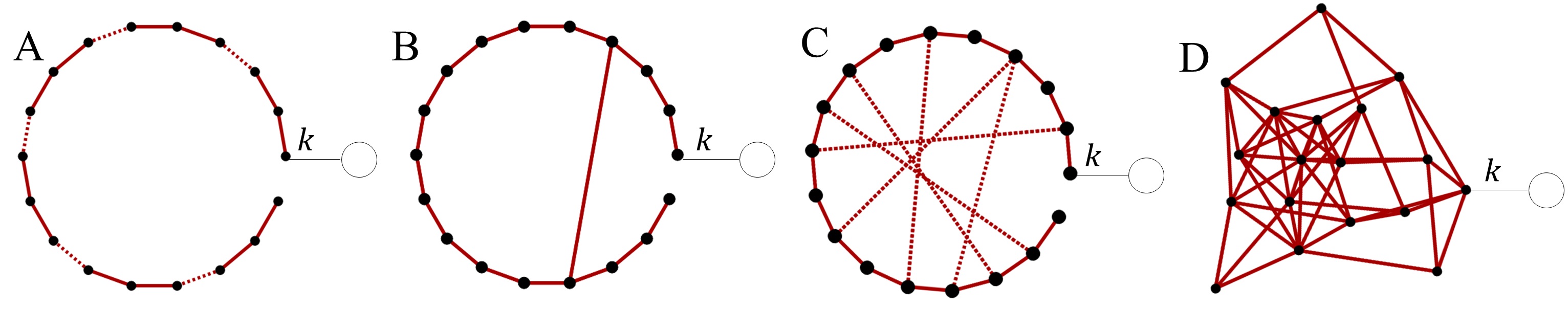}}
	\caption{\label{fig1}
		(Color online) Schematics of the network configurations. The white circle is the probe, coupled to a single node with a coupling strength $k$. Couplings between network oscillators are either strong (solid line) or weak (dashed line). Example (A) is a periodical chain where every third coupling strength is weaker. Example (B) is a homogeneous chain with one short-cut, with all coupling strengths equal. Example (C) is a small-world network constructed from a chain by adding several weak shortcuts and example (D) is an Erd\H{o}s-R\'{e}nyi random network.
	}
\end{figure}
%%%%%%%%%%%%%%%%%%%%%%%%%%%%%%%%%%%%%%%%

The reduced dynamics can be described exactly by a generalized quantum Langevin equation. In particular, the dissipation and memory effects are accounted for by the damping kernel $\gamma(t)=\sum\nolimits_{i=1}^N (k^2g_{i}^2/\Omega_{i}^2)\cos(\Omega_it)$. The spectral density of the environmental couplings is defined as

%%%%%%%%%%%% EQUATION 1 %%%%%%%%%%%%%%%%%%
\begin{equation}
\label{eq.1}
J(\omega)=\dfrac{\pi}{2}\sum\limits_{i} \dfrac{k^2g_{i}^2}{\Omega_i}\delta(\omega-\Omega_i).
\end{equation}
%%%%%%%%%%%%%%%%%%%%%%%%%%%%%%%%%%%%%%%%
 
The spectral density and the damping kernel are related as

%%%%%%%%%%%% EQUATION 2 %%%%%%%%%%%%%%%%%%
\begin{equation}
\label{eq.2}
J(\omega)=\omega\int^{t_{max}}_0 \gamma(t)\cos(\omega t)\, \mathrm{d}t\,.
\end{equation}
%%%%%%%%%%%%%%%%%%%%%%%%%%%%%%%%%%%%%%%%

For finite networks, the interaction time between the probe and the network controls the crossover between smooth and discrete spectral densities. For $t_{max}$ in eq.~(\ref{eq.2}) smaller than a characteristic recurrence time $\tau_f$, proportional to the size of the network, the spectral density will be a smooth function of frequency. In this regime, finite size effects are avoided and reduced dynamics correspond to a probe interacting with a continuum of frequencies, similar to that calculated from eq.~(\ref{eq.1}) in the limit $N\rightarrow\infty$. In $t_{max}>\tau_f$ regime, finite size effects start to play a prominent role.

In what follows, eq.~(\ref{eq.2}) will be used to calculate $J(\omega)$ throughout. We will work in the continuum regime when probing and engineering the spectral density, but it will be seen that when probing the topology, one must work in the discrete regime instead.

%%%%%%%%%%%% FIGURE 2 %%%%%%%%%%%%%%%%%%
\begin{figure}[t]
                \includegraphics[trim=0.3cm 0.1cm 0.3cm 0cm,clip=true,width=0.95\textwidth]{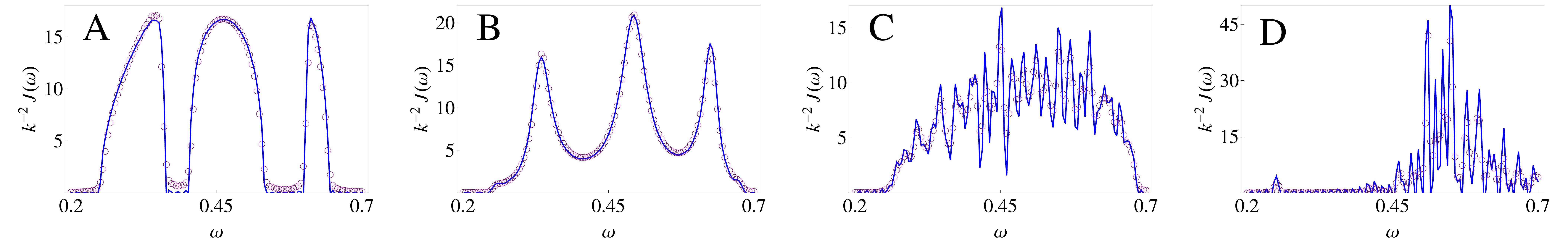}
            \caption{\label{fig2}
(color online) The spectral densities (solid line) of the configurations presented in Fig. \ref{fig1} for a system-network dimensionless coupling strength $k=0.01$. All networks have $N=200$ nodes with a frequency $\omega_0=0.25$. The spectral densities are probed (circles) with a constant interaction time $t=500$. Each circle corresponds to a different value of the system frequency, which is varied to sample the underlying $J(\omega)$. The system is initially in vacuum state and the network temperature is $T=5$. In the periodic chain (A), strong and weak couplings between network oscillators are $0.1$ and $0.06$, respectively. In the chain with a shortcut (B), all network couplings are set to $0.1$. In the small-world network (C), the strong and weak couplings are  $0.1$ and $0.003$, respectively. The random network (D) has equal couplings set to $0.05$.
}
      \end{figure}
%%%%%%%%%%%%%%%%%%%%%%%%%%%%%%%%%%%%%%%%

{\textbf{Engineering of the spectral density.}} We consider several different network configurations and calculate the respective spectral densities. By going beyond simple chain networks a very rich variety of different spectral densities arises. Schematics of network configurations considered in this work are presented in Fig.~\ref{fig1}, and the corresponding spectral densities are presented in Fig.~\ref{fig2}.

A simple homogeneous chain with springlike couplings has a single band. The lowest frequency where $J(\omega)$ is non-vanishing coincides with $\omega_0$ and the highest frequency increases as the stiffness of the chain is increased, while the magnitude is directly proportional to $k^2$. Band gaps can be created by periodically tuning some of the couplings in the chain to be weaker, which splits the single band to two or more bands \cite{dimerchainpaper}. We have checked that the number of bands coincides with the number of network oscillators in a strongly coupled group: an example is the trimer chain in Fig. \ref{fig1}A where several groups of three oscillators are weakly coupled as a chain, resulting in three bands as in Fig. \ref{fig2}A. A bigger difference between strong and weak network couplings leads to wider bandgaps.

Instead of tuning many couplings periodically, a similar effect can be achieved by adding just one additional link, or a shortcut, to a homogeneous chain, as in Fig. ~\ref{fig2}B. This leads to a structured $J(\omega)$ with spikes. The amount of spikes coincides with the number of oscillators before the shortcut, while the depth of the pseudogaps between the spikes depends on the coupling strength of the shortcut with stronger shortcut leading to deeper pseudogaps. An example with three spikes is shown in Fig.~\ref{fig2}B.

We also considered a small-world network and a random network. The former has a single band with a lot of fine structure. On the other hand, the random network produces highly structured spectral densities that can change completely from node to node. In the example presented in Fig.~\ref{fig2}D, the probe is coupled to a node with high connectivity, resulting in a $J(\omega)$ with a wide frequency range. Coupling to a node with low connectivity instead might result in a very narrow $J(\omega)$.

The ability to controllably engineer a wide variety of spectral densities highlights that complex quantum networks are ideal testbeds for fundamental studies on open quantum systems where, especially in the non-Markovian regime, a number of fundamental questions remain still unanswered.

{\textbf{Probing of the spectral density.}} We consider the case of a network with an unknown configuration but known temperature and show that the local value of the spectral density can be probed by measuring the expectation value of the probe number operator $\braket{n}{}=\braket{a^\dagger a}{}$, where $a^\dagger$ and $a$ are the system creation and annihilation operators. This is proportional to the average energy of the probe. It has been experimentally measured for specific systems such as trapped ions; see, e.g., \cite{ionreview} for a review. In the examples, we calculate the evolution of $\braket{n}{}$ by diagonalizing the network, solving the Heisenberg equations of motion in eigenmode picture, and returning to old variables.

Provided that the network size is not too small, one can easily check by comparison with the exact numerics that, for weak probe-network couplings and for times greater than the initial non-Markovian time in the continuum regime $t_{max}<\tau_f$, the mean value of the probe number operator is well approximated by the expression  $\braket{n(t)}{}=e^{-\Gamma t}\braket{n(0)}{}+N(\omega_S)(1-e^{-\Gamma t})$, with $\Gamma=J(\omega_S)/\omega_S$ and where $N(\omega_S)=(e^{\omega_S/T}-1)^{-1}$ is the thermal average boson number, $T$ being the temperature of the environment, in units of the Boltzman constant $k_B$ \cite{energyeq}. The spectral density can be obtained by probing the network scanning over different frequencies $\omega_S$ via the formula

%%%%%%%%%%%% EQUATION 3 %%%%%%%%%%%%%%%%%%
\begin{equation}
\label{eq.3}
J(\omega_S) = \dfrac{\omega_S}{t} \ln \left(\dfrac{\Delta n (0)}{ \Delta n(t)}\right),
\end{equation}
%%%%%%%%%%%%%%%%%%%%%%%%%%%%%%%%%%%%%%%%
\noindent
where $\Delta n (t) =  N(\omega_S) - \langle n (t) \rangle$ and $t$ a suitable interaction time. We stress once more that $t$ can be chosen within a wide interval, i.e. it should be in the continuum regime and longer than the initial non-Markovian time. A rough estimate of the non-Markovian correlation time $\tau_R$ is given by $\omega_{max} \tau_R \gg 1$, where $\omega_{max}$ is the largest frequency where the spectral density is non-vanishing: for weak couplings it is typically very short.

%%%%%%%%%%%% FIGURE 3 %%%%%%%%%%%%%%%%%%
\begin{figure}[t]
	\includegraphics[trim=0cm 0cm 0.18cm
 0cm,clip=true,width=0.95\textwidth]{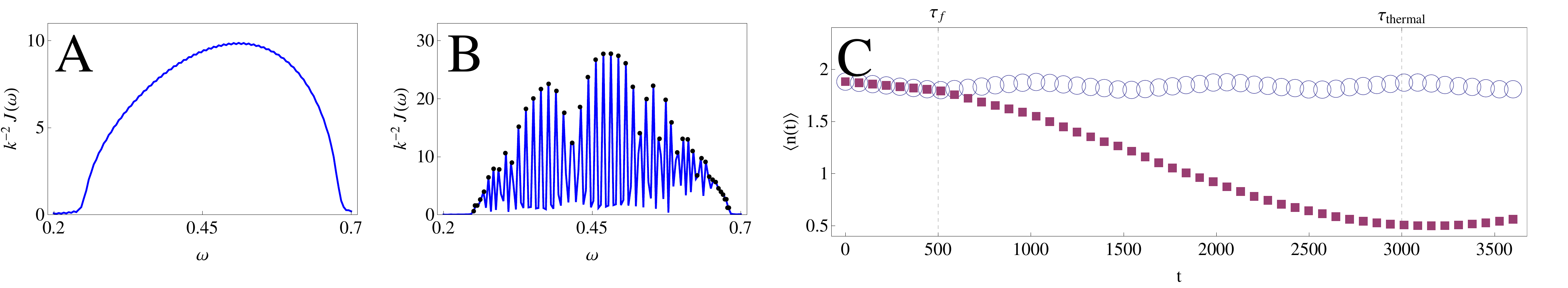}
	\caption{\label{fig3}
		 (color online) An example of eigenfrequency probing with a homogeneous chain. In (A), the detected spectral density is shown for $t\approx0.9 \tau_f$, where $t$ is the interaction time and $\tau_f$ the recurrence time. The result is a smooth function of frequency. In (B), detected spectral density is shown for $t\approx3 \tau_f$. The discrete spectrum emerges and reveals the eigenfrequencies marked by dots. In (C), the evolution of $\langle n(t) \rangle$ is compared for $\omega_S$ at an eigenfrequency (squares) and just $2 \%$ off (circles). The difference is clear and causes the emergence of discrete spectrum in (B). Also depicted is the time  $\tau_{thermal}$, after which the energy flow is reversed even for a system at an eigenfrequency. The chain, initially in vacuum state, was 50 oscillators long, with bare frequency $\omega_0=0.25$ and coupling constant $g=0.1$, with a system-network coupling strength $ k=0.0025 $. The probe was initially in squeezed vacuum with squeezing parameter $r=1$ and phase space angle $\phi=\pi/2$.}
	
\end{figure}
%%%%%%%%%%%%%%%%%%%%%%%%%%%%%%%%%%%%%%%%  

In Fig.~\ref{fig2} we test the effectiveness of the probing scheme for the spectral densities of the networks presented in Fig.~\ref{fig1} by calculating the r.h.s. of eq.~(\ref{eq.3}) with exact numerics for many different values of $\omega_S$. All quantities, namely coupling strengths, times and temperatures, are given in arbitrary units as referred to a fixed frequency unit. The probing scheme works well for all four examples. In particular it is seen that while the rich structure of Figures \ref{fig1}C and \ref{fig1}D would require dense sampling to resolve, the samples themselves are accurate. Beyond the examples presented here, we checked the accuracy of the probing for a wide range of spectral densities and parameter ranges and found similar results. We also tested probing with different states of same energy, finding no differences.

{\textbf{Probing of the network topology.}} The spectral density alone does not allow one to determine all properties of the underlying quantum network. It can be shown \cite{linearmappingA, linearmappingB, linearmappingC} that any spectral density that can be produced by a diagonal Hamiltonian can be reproduced by a linear harmonic chain with nearest-neighbor couplings only. Since any network configuration, no matter how complex, can be presented in the eigenmode picture with a diagonal Hamiltonian, this means that the network shape is never unique to a given spectral density.

In the following we present a general approach allowing for the full characterisation of the network structure and topology. Remarkably, our detection method only relies on the use of a single quantum probe and on the measure of the mean excitation number $\langle n(t) \rangle$ at a certain time within a broad characteristic time interval. However, in order to retrieve the complete information on the network we need to (i) couple the probe to more than a single node; (ii) measure the excitation number in the discrete regime $t_{max}>\tau_f$. The latter point is related to the fact that short-time reduced dynamics are not determined by the entire network \cite{effectivenvironment}. This technique is, inevitably, much more expensive than the probing of the spectral density. However, to the best of our knowledge, it is the first demonstration that the full information on the network can be extracted by means of a single quantum probe.

We will assume to have some preliminary knowledge on the size of the network, i.e., on the number of nodes $N$. As explained when introducing the microscopic model, a generic oscillator network is determined by matrix \textbf{A}, which provides the oscillator frequencies as well as the structure of the network. We will show how to reconstruct both the matrix $\bf{D}$ and $\bf{K}$ from which matrix $\bf{A}$ can be obtained via $\textbf{A} = \textbf{K}\textbf{D}\textbf{K}^T $.

To construct $\bf{D}$, one needs to detect the eigenfrequencies $\Omega_i$, which can be done using an approach similar to $J(\omega)$ probing. As explained before, for sufficiently short interaction times, the detected $ J \left( \omega\right) $, given by eq.~(\ref{eq.3}), is a smooth function of frequency as if the network had a continuum of eigenfrequencies. With longer interaction times the discrete spectrum emerges. This is essentially caused by differences in the heat flow between the probe and the network in the long interaction time regime, leading to a significantly larger values of $\Delta n (t)$ when $\omega_S$ is close to an eigenfrequency. The sensitivity of the eigenfrequency detection depends on coupling strength $k$, with weaker values leading to more accurate results. We stress that this behaviour is universal to all finite networks. An example is shown in Fig. \ref{fig3}, which also demonstrates the accuracy of the eigenfrequency detection when $k$ is weak. Typically eigenfrequency detection is the most robust part of topology probing.

While $\Omega_i$ do not depend on the node the probe is coupled to, the probe must interact with all eigenmodes to detect all $\Omega_i$, which is not always the case. On the other hand, the couplings to eigenmodes can be changed by coupling the probe to different node(s), which allows one in principle to always detect all of them. If $\Omega_i$ are all distinct, as is the case for generic networks, knowing $N$ gives the number of $\Omega_i$, and consequently one knows when all eigenfrequencies have been found.

The elements of $\bf{K}$ are detected in two steps. First step is to detect the modulus of $\bf{K}$. The coupling strengths to eigenmodes are related to the rows of matrix \textbf{K} as $g_i(j)=K_{ji}$, with $g_i$ describing the coupling between the probe and the eigenmode $i$ and $j$ the index of the node the probe is coupled to. In other words, each node $j$ corresponds to a different $H_I$ and therefore, different set of coupling strengths. Thus, the elements $K_{ji}$ can be obtained by probing sequentially each node $j$ of the network at the eigenfrequencies $\Omega_i$, i.e. evaluating decay of $\langle n(t) \rangle$ at $ \omega_S=\Omega_i $, and using eq.~(\ref{eq.3}) to determine $J(\Omega_i)$. From eq.~(\ref{eq.1}), it can be seen that $\int^{\infty}_0 \frac{2}{\pi} J(\omega) \omega \mathrm{d}\omega\, = \sum_i k^2g_i^2$. Keeping in mind that we wish to have a value for each of the $g_i$, we approximate the integral with a finite sum using $J(\Omega_i)$ and $\Omega_i$. By identifying the terms in the summations on both sides and solving for $g_i$ we arrive at $|g_i | = \sqrt{2 J(\Omega_i) \Omega_i \Delta \Omega_i / (\pi k^2)}$, where $\Delta \Omega_i = \Omega_i - \Omega_{i+1}$ is the sampling interval.

Second step is to detect the signs of the elements of \textbf{K} by probing pairs of nodes in the same way as previously, e.g. nodes 1 and 2. In this way one detects $|K_{1i} + K_{2i} |$ (with probe frequency $\Omega_i$) and, by comparing it to the values obtained by the single node probing, namely $|K_{1i}| + |K_{2i} |$, one can reveal whether the elements have same or opposite sign. In principle, the choice for the tolerance value used here can create some ambiguity, but this can be done after probing is over and the final result remains mostly unchanged for a wide range of choices. Running over the eigenfrequencies $\Omega_i$ and running over $N-1$ pairs of nodes completes the relative signs of the complete matrix. The knowledge of relative signs is sufficient since the adjacency matrix $\bf{A} = \bf{KDK^T} $ is invariant under a global sign change of $\bf{K}$.

Even small errors in the previous steps easily destroy the orthogonality  of $\bf{K}$, leading to large errors in the reconstruction to the overall network structure and topology. To counter this, the orthogonality needs to be enforced. This is straightforward to do since it can be shown \cite{nearestorthogonalmatrix}  that any matrix \textbf{M} which has a singular value decomposition $ \textbf{M}=\textbf{W} \textbf{N}\textbf{V}^{\dagger} $ has a unique nearest orthonormal matrix, and it is $ \textbf{WV}^{\dagger} $.

\begin{figure}[t]
                \includegraphics[trim=0cm 0cm 0cm 0cm,clip=true,width=0.7\textwidth]{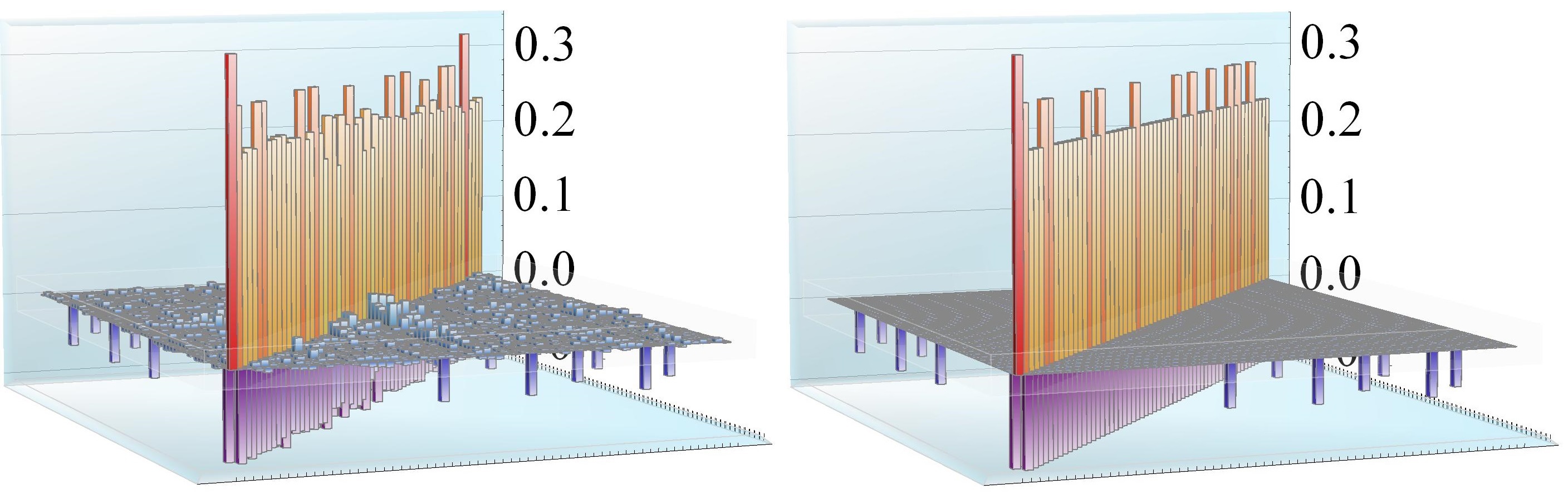}
            \caption{\label{fig4}
(color online) The detected matrix $\bf{A}$ (left), compared to the original one (right) for a smallworld network of $ 60 $ nodes. The initial state for the probe is squeezed vacuum with squeezing parameter $r=1$ and phase space angle $\phi=\pi/2$, while the network is initially in vacuum state. The chain couplings have a coupling strength of 0.2 while the seven extra links have a coupling strength of 0.1. Bare frequencies are $ \omega_0=0.25 $.
}
      \end{figure}

To summarize, one first needs to detect the eigenfrequencies $\Omega_i$. This determines matrix $\bf{D}$. Then, all $N$ nodes are probed at eigenfrequencies individually, followed by the probing of $N-1$ pairs of nodes, which determines matrix $\bf{K}$ up to a global sign which can be arbitrary. The nearest orthonormal matrix to detected $\bf{K}$ is calculated and used to calculate the adjacency matrix from $\bf{A} = \bf{KDK^T} $. The method above can be used for a full reconstruction of a large oscillator network of an unknown topology. As an example we show in Fig.~\ref{fig4} a comparison between the original and reconstructed adjacency matrices of a small-world network similar to that of Fig.~\ref{fig1}C. The accuracy of the reconstruction depends mostly on the magnitude of the adjacency matrix elements and the used interaction time $t$: accuracy improves for larger values of matrix elements and using an interaction time in the range $\tau_f<t<\tau_{thermal}$, where $\tau_{thermal}$ is the time before energy flow is reversed for a system at an eigenfrequency, see Fig.~\ref{fig3}C. In particular, links that are very weak relative to others can be missed, while a network where all couplings are strong leads to best results. The biggest drawback of the method is that the amount of measurements one needs to do scales with $N^2$. This is a consequence of having an unknown topology with unknown and possibly distinct coupling strengths. Whether it is possible to do better without making additional assumptions about the network is still very much an open question.

%%%%%%%%%%%% FIGURE 4 %%%%%%%%%%%%%%%%%%
%%%%%%%%%%%%%%%%%%%%%%%%%%%%%%%%%%%%%%%% 

\section*{Discussion}

We have explored bosonic complex networks from two complementary perspectives. We have shown how, having access and control of the network structure, we can engineer a wide variety of  {\it ad hoc} structured environments. We have demonstrated that even small changes in the connectivity of the network, such as the introduction of an additional link, may cause a dramatic change in the shape of the spectral density. Highly structured spectral densities are found, e.g., in biological systems and, very recently, approaches of open quantum systems theory have been used to understand the loss of coherence in these systems \cite{quantbio1,quantbio2}. As biological systems are very difficult to test in the laboratories, it is highly desirable to develop ways to simulate their dynamics in a controllable way. Possible experimental implementations of the model considered here include optical modes in cavities \cite{opticalmodes}, ions in segmented traps \cite{segmentedtraps}, mechanical resonator arrays \cite{resonatorarrays}, and cluster states created from an ultrafast frequency comb \cite{clusterstates}. While implementing an arbitrary oscillator network is still beyond the reach of experimentalists, simple networks have been implemented with all three platforms and the realization of more complex networks is under active research.

With a shift in perspective to a complementary view point, we have investigated the possibility of revealing unknown properties of a complex quantum network by means of a single quantum probe suitably coupled to the network. We have developed and illustrated a method for efficient detection not only of the spectral density but, in principle, of the entire network structure, showing its effectiveness for virtually any network configuration. Our results are the first demonstration of the power of quantum probing for complex quantum networks going beyond toy models consisting of tens of nodes. While the present results are still a proof-of-principle demonstration, due to the generality of the system considered, they pave the way to the study of quantum probing in a variety of  physical scenarios.

\section*{Acknowledgements}

This work is supported by the Horizon 2020 EU Collaborative project Quantum Probes for Complex Systems (Grant Agreement 641277), by project FIS2014–60343-P (NOMAQ), and by the Academy of Finland (Project no. 287750).
Financial support from Jenny and Antti Wihuri foundation and the University of Turku doctoral programme in Physical and Chemical sciences is also acknowledged. J.N and J.P. thank IFISC for the hospitality. F.G. acknowledges support from the post-doctoral programme of UIB.

\section*{Author contributions statement}

J.N., S.M. and J.P. planned the research. J.N. did most calculations and numerical simulations, with help from F.G. and R.Z.. All authors contributed in analysing and discussing results and writing the manuscript.

\section*{Additional information}

\textbf{Competing financial interests} The authors declare no competing financial interests.

\end{document}